\documentstyle[epsfig,mathptm]{mn} 
\newif\ifAMStwofonts
\AMStwofontstrue               
\DeclareMathVersion{bold}             
\DeclareMathAlphabet{\mathbfit}{OT1}{cmr}{bx}{it}
\SetMathAlphabet\mathbfit{bold}{OT1}{cmr}{bx}{it}
\DeclareMathAlphabet{\mathbfss}{OT1}{cmss}{bx}{n}
\SetMathAlphabet\mathbfss{bold}{OT1}{cmss}{bx}{n}
\ifAMStwofonts
  \ifCUPmtlplainloaded \else
    \DeclareSymbolFont{UPM}{U}{eur}{m}{n}
    \SetSymbolFont{UPM}{bold}{U}{eur}{b}{n}
    \DeclareSymbolFont{AMSa}{U}{msa}{m}{n}
    \DeclareMathSymbol{\upi}{0}{UPM}{"19}
    \DeclareMathSymbol{\umu}{0}{UPM}{"16}
    \DeclareMathSymbol{\upartial}{0}{UPM}{"40}
    \DeclareMathSymbol{\leqslant}{3}{AMSa}{"36}
    \DeclareMathSymbol{\geqslant}{3}{AMSa}{"3E}

    \let\leq=\leqslant 

  \fi
\fi

\newcommand{\apjs}{{ApJS}}
\newcommand{\apjl}{{ApJL}}
\newcommand{\aj}{{ AJ}}

\newcommand{\mnras}{{MNRAS}}

\newcommand{\apj}{{ ApJ}}

\title
[Velocities, CMB and Supernovae]
{Cosmological Parameters from Velocities, CMB and Supernovae} 
\author[S.L.~Bridle et al.]
{S.L.~Bridle$^1$, I.~Zehavi$^2$, A.~Dekel$^3$, O.~Lahav$^{4,3}$, 
M.P.~Hobson$^1$, A.N.~Lasenby$^1$\\
$^1$Astrophysics Group, Cavendish Laboratory,  Madingley Road, 
Cambridge CB3 0HE, UK\\
$^2$NASA/Fermilab Astrophysics Group, Fermi national Accelerator Laboratory,
Box 500, Batavia, IL 60510-0500, USA\\
$^3$Racah Institute of Physics, The Hebrew University, Jerusalem 
91904, Israel\\
$^4$Institute of Astronomy, Madingley Road, Cambridge CB3 0HA, UK\\
 }

\date{}
\pagerange{\pageref{firstpage}--\pageref{lastpage}}
\pubyear{2000}
\begin{document}
\maketitle
\label{firstpage}

\begin{abstract} 
We compare and combine likelihood functions of the cosmological
parameters $\Omega_{\rm m}$, $h$ and $\sigma_8$, from peculiar
velocities, CMB and type Ia supernovae. These three data sets
directly probe the mass in the Universe, without the need
to relate the galaxy distribution to the underlying mass
via a ``biasing'' relation. We include the recent results from the 
CMB experiments BOOMERANG and MAXIMA-1. Our analysis assumes a flat 
$\Lambda$CDM cosmology with a scale-invariant adiabatic initial power 
spectrum and baryonic fraction as inferred from big-bang nucleosynthesis.
We find that all three data sets agree well,
overlapping significantly at the 2$\sigma$ level. This therefore
justifies a joint analysis, in which we find a joint best
fit point and $95\%$ confidence limits of $\Omega_{\rm m}=0.28\,
(0.17,0.39)$, $h=0.74\, (0.64,0.86)$, and $\sigma_8=1.17\,
(0.98,1.37)$. In terms of the natural parameter combinations for these
data $\sigma_8\Omega_{\rm m}^{0.6}=0.54\, (0.40,0.73)$, $\Omega_{\rm
m} h = 0.21\, (0.16,0.27)$. Also for the best fit point,
$Q_{\rm{rms-ps}} = 19.7\mu$K and the age of the universe is $13.2$
Gyr.
\end{abstract}

\section{Introduction}
\label{intro}

A simultaneous analysis of the constraints placed on the cosmological
parameters by various different kinds of data is essential because
each different probe typically constrains a different combination of
the parameters. By considering these constraints together, one can
overcome any intrinsic degeneracies to estimate each
fundamental parameter and its corresponding random uncertainty. The
comparison of constraints can also provide a test for the validity of
the assumed cosmological model or, alternatively, a revised evaluation
of the systematic errors in one or all of the data sets. 
Recent papers that combine information from several data sets simultaneously 
include Gawiser \& Silk (1998), Bridle et al. (1999), Bahcall et al. (1999), 
Bond \& Jaffe (1999), Lineweaver (1998) and Lange et al. (2000).

Galaxy motions relative to the Hubble flow arise from the gravitational 
forces due to mass-density fluctuations; they therefore
reflect the underlying distribution of matter (both dark and luminous), 
and can thus provide constraints on the cosmological density parameter
$\Omega_{\rm m}$ and the fluctuation amplitude parameter
$\sigma_8$. For example, constraints on the cosmological parameters
were obtained by Zaroubi et al. (1997) and Freudling et al. (1999) 
from a likelihood analysis of the Mark III and
SFI catalogues of peculiar velocities, in the framework of
COBE-normalised cold dark matter (CDM) models and Gaussian
fluctuations and errors. 
The anisotropies in the cosmic microwave background (CMB) depend on
the state of the universe at the epoch of recombination, on 
the global geometry of space-time and on any re-ionization. 
Thus they provide a powerful and potentially accurate 
probe of the cosmological parameters (see Hu, Sugiyama and Silk 1997
for a review). 
With the recent release of results from a new generation of CMB
experiments BOOMERANG and MAXIMA-1 have come a number of parameter estimation
analyses, including those by Lange et al. (2000), Balbi et al. (2000) and
Tegmark \& Zaldarriaga (2000).
The distances of type Ia supernovae (SN) can now be measured at large
redshift. Thus they can provide constraints on the acceleration of the 
universal expansion, and the
corresponding parameters $\Omega_{\rm m}$ and $\Omega_{\Lambda}$, via a
classical cosmological test based on the luminosity-redshift
relation. These three cosmic probes allow direct dynamical constraints
free of assumptions regarding the ``biasing" relation between the
distribution of galaxies and the underlying matter density, 
which are unavoidable when interpreting galaxy redshift surveys. 

Various analyses have been performed in which pairs of these data sets
are used to place constraints in the $\Omega_{\rm m}$,
$\Omega_{\Lambda}$ plane: Efstathiou et al. (1999) 
investigate CMB and SN; Zehavi \& Dekel (1999) 
explore peculiar velocities with SN. In this
work, we perform a joint analysis of all three data sets. We restrict
the analysis to the scale-invariant flat $\Lambda$CDM model, which is
motivated by theoretical arguments based on the inflation
scenario, and is consistent with CMB observations (e.g. Bond \&
Jaffe 1998 find $n\sim 1$ from the COBE data; recent analyses of the
BOOMERANG and MAXIMA-1 data find $\Omega_{\rm m}+\Omega_{\Lambda}\sim 1$).
In addition we use the nucleosynthesis constraint of $\Omega_{\rm b}
h^2 \sim 0.019$ (Tytler et al. 2000), although we discuss the
validity of this assumption in the light of the recent measurements of
the CMB second peak height. 

An earlier paper (Bridle et al. 1999) investigated the combination of
constraints from CMB data, the abundance of clusters of
galaxies (Eke et al. 1998) and the IRAS 1.2 Jy redshift survey
(Fisher, Scharf \& Lahav 1994). These data sets were found to be in
good agreement, once the densities of galaxies and mass are assumed 
to be related via a linear biasing parameter. In this paper we focus 
on the implications of combining three dynamical data sets that are 
free of galaxy-biasing uncertainties.
In \S~\ref{sdata} we introduce each of the data sets and outline the 
theory used to link the constraints from each and in \S~\ref{results} 
we compare and combine the constraints from the different data.

\section{The three probes of mass} 
\label{sdata}

\subsection{Peculiar Velocities}
\label{pv}

We consider two catalogues of galaxy peculiar velocities (PV).
One, Mark III (Willick et al. 1997), contains $\sim\!3000$ galaxies
within $\sim\!70 h^{-1}$ Mpc 
($h\equiv H_0/(100$~\mbox{$\rm{km} \rm{s}^{-1} \rm{Mpc}^{-1}$}$)$). 
The other, SFI (Haynes et al. 1999a,b),
consists of $\sim\!1300$ spiral galaxies but with a more uniform
spatial coverage in a similar volume. The error per galaxy is 
$15-21\%$ of its distance. The constraints obtained from the two data 
sets were found to be very similar (e.g Zehavi \& Dekel 2000). We 
therefore choose to perform our analysis here on the SFI catalogue. 

The analysis follows in general the maximum-likelihood method of 
Zaroubi et al. (1997) and Freudling et al. (1999). 
The density and velocity fluctuations are assumed to be a random
realization of a Gaussian field and to obey the linear approximation 
to gravitational instability (with one caveat discussed below). 
The likelihood of a given set of values for the cosmological parameters 
of interest is calculated by comparing the observed velocity correlations 
with those predicted by theory based on this set of parameters, under the 
assumption that the errors in the observed velocities are Gaussian. 

The CDM power spectrum form used in the likelihood analysis 
has a shape parameter $\Gamma$, as provided by Sugiyama (1995, Eq. [3.9]), 
which determines the wavenumber at the
peak of the power spectrum $P(k)$ in terms of $h$, $\Omega_{\rm{m}}$ 
and the baryonic content of the universe $\Omega_{\rm{b}}$. 
This is independent of $\Omega_{\Lambda}$, and we assume that the
power spectrum is initially scale-invariant ($n=1$).  
For the baryonic content we adopt the value favored by Deuterium
abundance in the context of Big-Bang nucleosynthesis (Tytler et
al. 2000) $\Omega_{\rm{b}} h^2=0.019$.
Note that while Freudling et al. (1999) used COBE-normalised models, we 
perform our current analysis of peculiar velocities with the amplitude of 
fluctuations as a free parameter. The COBE constraint enters the joint
analysis later as part of the independent CMB data set. 
In this paper we thus choose as our free fundamental parameters 
the dimensionless Hubble constant $h$,
the total matter density, $\Omega_{\rm m}$ ($=1-\Omega_{\Lambda}$ here),
and the normalisation mass-density parameter $\sigma_8$.

Note that the linear analysis of the velocity data addresses the scaled 
power spectrum $P(k) \Omega_{\rm m}^{1.2}$ rather than $P(k)$ itself, 
and it therefore 
constrains the combination of parameters $\sigma_8 \Omega_{\rm m}^{0.6}$,
which serves as a measure of the power-spectrum amplitude. 
This result is almost independent of $\Omega_{\Lambda}$ (Lahav et al. 1991). 
Its shape is controlled by another combination $\Gamma\sim \Omega_{\rm m} h$. 
These combinations are therefore the natural parameters for the velocity 
analysis. The wavenumber range covered is roughly 
$0.05<k\,\, (h \,\,{\rm Mpc}^{-1})<0.2$. 

In order to account for nonlinear effects acting on small scales, we 
add to the linear velocity correlation model an additional 
free parameter, $\sigma_{\rm v}$, representing an uncorrelated velocity 
dispersion at zero lag.  This is a simple way to model 
small-scale random virial motions, but it can also be interpreted as 
an addition to the errors that enter the likelihood analysis. 
The parameter $\sigma_{\rm v}$ is allowed to vary together
with the other model parameters.
By this procedure, the cosmological parameters of interest are
properly determined by the linear part of the fluctuations on large scales, 
while the undesired nonlinear effects are detached and ``absorbed" 
by the additional free parameter.
This procedure has been explored already in Freudling et al. (1999, \S6.3.2),
who obtained for the SFI catalogue a best-fit value of $\sigma_{\rm
v}=200 {\rm km s^{-1}}$, resulting in values of 
$\sigma_8 \Omega_{\rm m}^{0.6}$ lower by $10-20\%$
than the values obtained without this additional term.
This and other ways of correcting for nonlinear effects have been 
recently found to provide consistent and more reliable results, 
based on improved mock catalogues drawn from high-resolution simulations 
and a Principal Component analysis (Silberman et al., in preparation).
The results obtained using this nonlinear correction serve as the
back-bone for the joint analysis and the figures of the current paper, 
where we marginalise over $\sigma_{\rm v}$.
By marginalisation we mean integrating likelihoods over a fixed range
with a uniform prior. In this case the range is 
$0$~km~s$^{-1} <\sigma_{\rm v}<400$ km s$^{-1}$, 
which we have chosen to be large enough so that our results are not
affected by the exact values of these limits. 
The results when we set $\sigma_{\rm v}$ to its (PV) best-fit value of $200$
km s$^{-1}$ are quite similar. For comparison we also refer to the
results obtained without the nonlinear correction.

\begin{figure*}
\centerline{
\mbox{\begin{tabular}{ccc}
(a)$\qquad$$\qquad$$\qquad$$\qquad$$\qquad$$\qquad$$\qquad$$\qquad$$\qquad$&
(b)$\qquad$$\qquad$$\qquad$$\qquad$$\qquad$$\qquad$$\qquad$$\qquad$$\qquad$&
(c)$\qquad$$\qquad$$\qquad$$\qquad$$\qquad$$\qquad$$\qquad$$\qquad$
\end{tabular}}}
\centerline{
\begin{tabular}{ccc}
\mbox{\epsfig{file=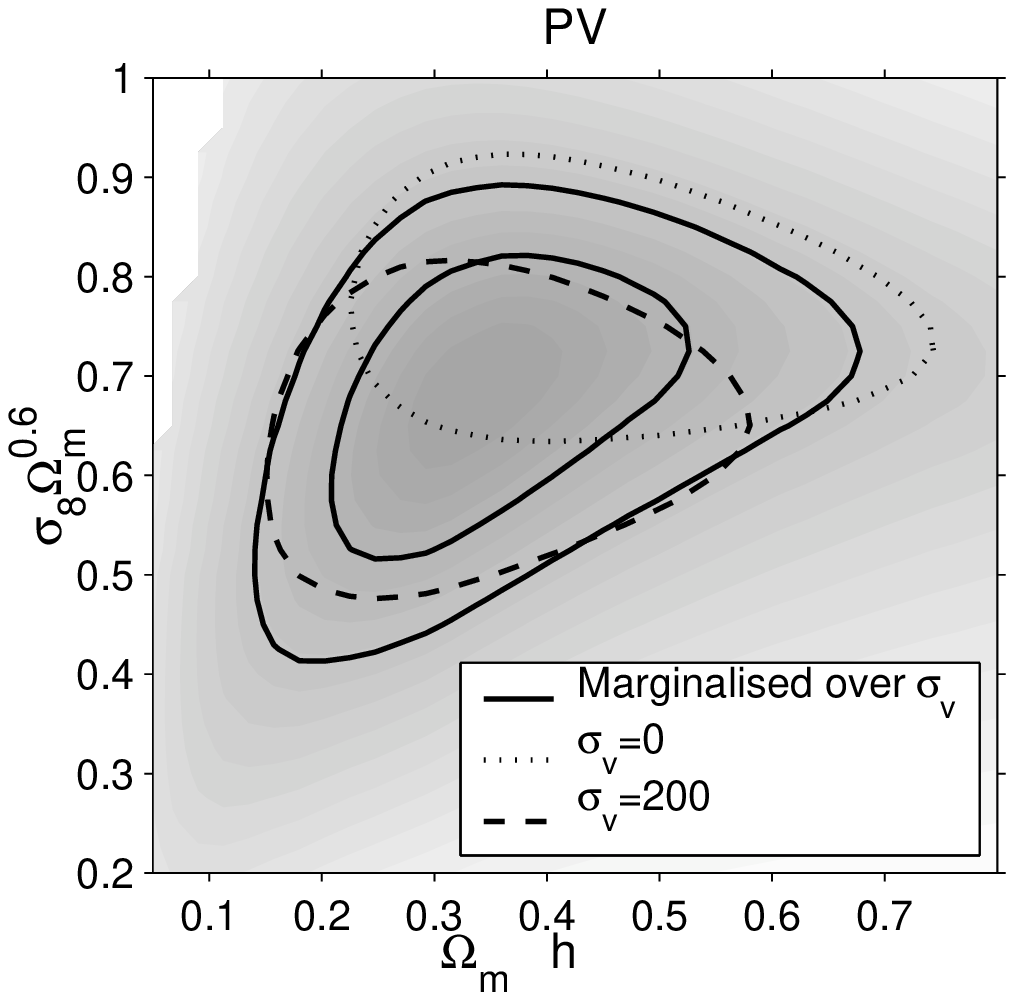,height=5.5cm,angle=90}}&
\mbox{\epsfig{file=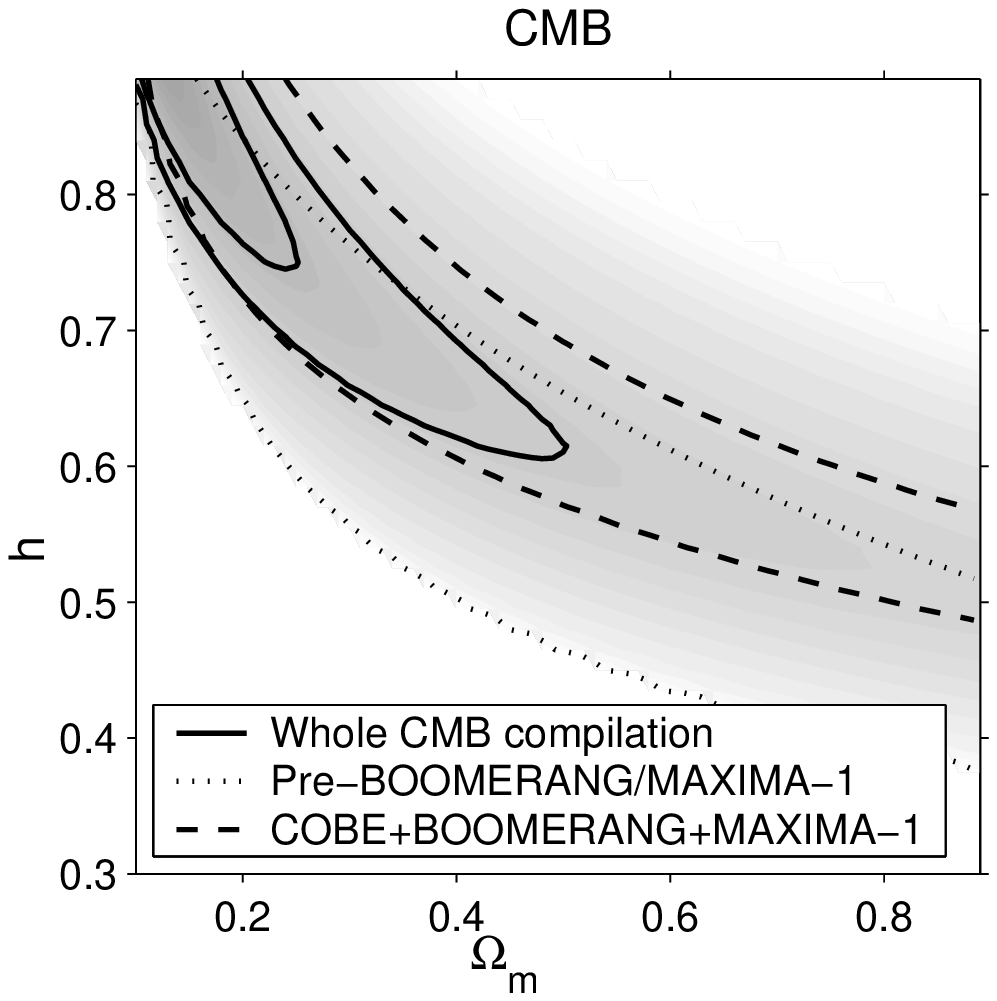,height=5.5cm,angle=90}}&
\mbox{\epsfig{file=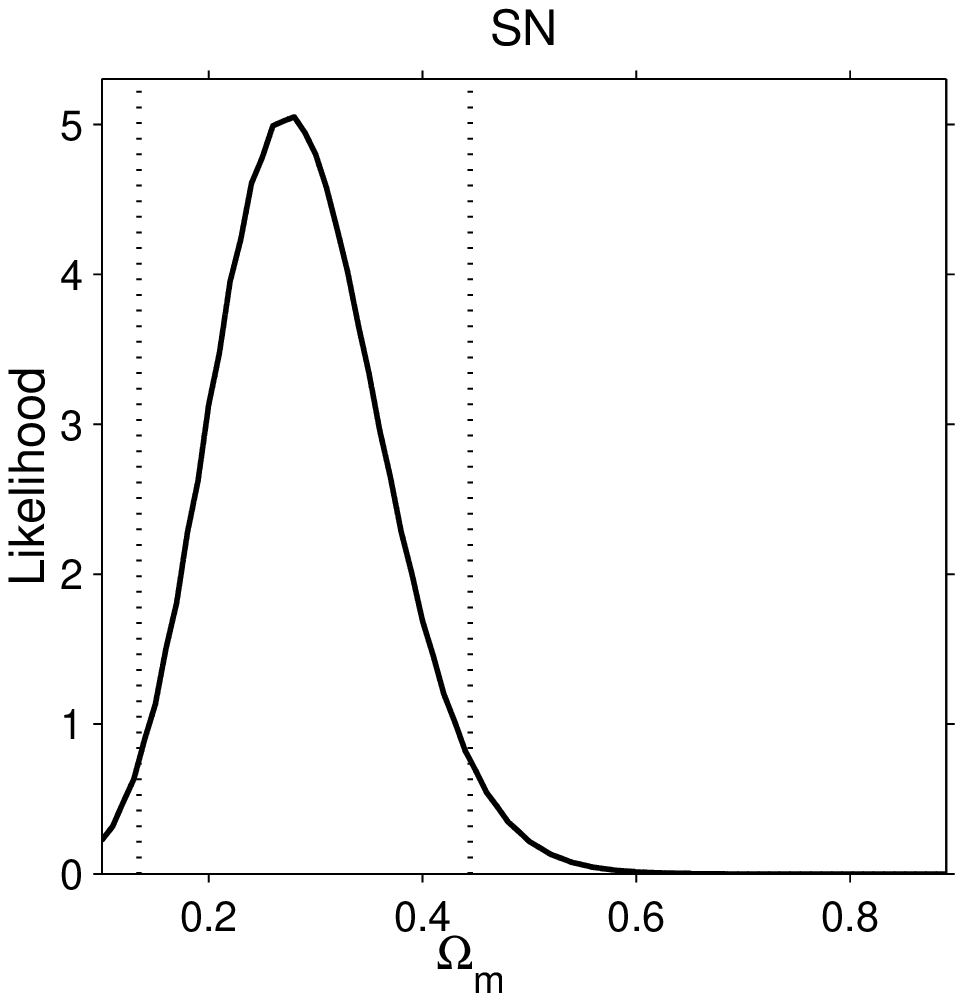,height=5.35cm,angle=90}}
\end{tabular}
}
\caption{(a) The constraints from the peculiar velocity data plotted in the 
$\Omega_{\rm m} h$, $\sigma_8\Omega_{\rm m}^{0.6}$ plane (marginalised
over $\Omega_{\rm m}$). The solid lines show $68$ and $95$ per cent
confidence limits after marginalisation over $\sigma_{\rm v}$, which is used
for the main part of the analysis. The dotted line shows the $95$ per
cent constraint if the parameter $\sigma_{\rm v}$ is not used (or
equivalently, $\sigma_{\rm v}=0$) the effect of which on the results of the 
joint analysis is discussed in section \ref{results}. 
The result of setting  $\sigma_{\rm v}=200$ km s$^{-1}$, the best fit
value to PV, is also shown (dashed line is the $95$ per cent contour). 
(b) The constraints in the $\Omega_{\rm m}$, $h$ plane from the
CMB data (marginalised over $\sigma_8$). 
The solid lines show the $68$ and $95$ per cent limits using the whole
CMB data compilation, which is used for that main results of this
paper. 
The $95$ per cent contours from the pre-BOOMERANG (Antarctica
flight)/MAXIMA-1 data (pre-BM) and
from just COBE$+$BOOMERANG (Antarctica flight)$+$MAXIMA-1 data are
shown by the dotted and dashed lines respectively.
(c) The supernova constraint on $\Omega_{\rm m}$; the dotted line
shows the 95 per cent confidence limits.
}
\label{data}
\end{figure*}

Fig.~\ref{data} (a) shows the two-dimensional probability distribution
for the PV data alone in the parameter space ($\sigma_8 \Omega_{\rm m}^{0.6}$,
$\Omega_{\rm m} h$)
(the constraints in the $\sigma_8 \Omega_{\rm m}^{0.6}$, $\Omega_{\rm
m} h$ plane are virtually insensitive to the value of $\Omega_{\rm
m}$, but for definiteness we have marginalised over $\Omega_{\rm m}$).
The velocity data constraints at the $95\%$ confidence level are
$0.48<\sigma_8 \Omega_{\rm m}^{0.6} <0.86$ and 
$0.16<\Omega_{\rm m} h< 0.58$, with roughly uncorrelated errors. 
For comparison, without
the $\sigma_{\rm v}$ term the results are 
$0.65<\sigma_8 \Omega_{\rm m}^{0.6} <0.89$ and $0.25<\Omega_{\rm m} h< 0.66$.

\subsection{The Cosmic Microwave Background}
\label{cmb}

\begin{figure}
\centerline{\vbox{
\epsfig{file=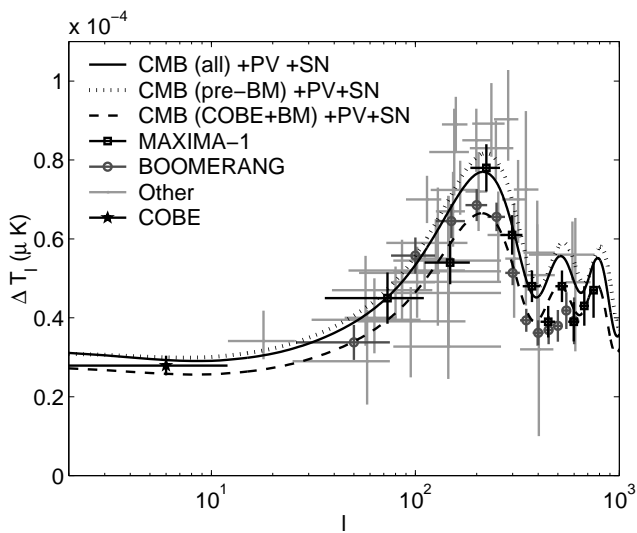,width=8.5cm}
}}
\caption{ The CMB data used. The solid line is the theoretical power
spectrum for the best fit point to the PV, SN and CMB when the whole
CMB data compilation is used ($h=0.74$, $\Omega_{\rm m}=0.28$,
$\sigma_8=1.16$).  
The dotted line is the best fit to PV$+$CMB$+$SN when just the 
pre-BOOMERANG (Antarctica flight)/MAXIMA-1 CMB data (pre-BM) is used
($h=0.65$, $\Omega_{\rm m}=0.34$, $\sigma_8=1.09$). 
The dashed line is from when just
COBE, BOOMERANG (Antarctica flight) and
MAXIMA-1 are used in the CMB compilation ($h=0.75$, $\Omega_{\rm
m}=0.29$, $\sigma_8=1.09$).
\label{cmbdata}}
\end{figure}

We use the same compilation of CMB anisotropy measurements as in
Bridle et al. (1999), supplemented by the new TOCO points (Miller et
al. 1999), the BOOMERANG North American test flight results (Mauskopf
et al. 1999), 
the BOOMERANG $10$ day Antarctica flight (de Bernardis
et al. 2000) and the MAXIMA-1 results (Hanany et al. 2000).
Since window functions are not yet available for these last three
experiments we assume, for each band power estimate, 
Gaussian window functions which fall by a
factor of $1/e$ at $\ell_{\rm{min}}$ and $\ell_{\rm{max}}$
as specified in de Bernardis et al. (2000) and Hanany et al. (2000). 
We also marginalise over the $10$ and $4$ per cent
calibration uncertainties quoted, respectively, for the BOOMERANG
Antarctica and MAXIMA-1 results,
fully taking into account the correlated nature of the calibration 
errors (Bridle et al., in preparation).
The full compilation is plotted in Fig.~\ref{cmbdata}. 
We compute the likelihood of the angular power spectra using the
flat-band power method (e.g. Hancock et al. 1998). In addition to the
assumptions already listed in the previous section, we assume there is
negligible re-ionization and that there are negligible tensor
contributions, as predicted by most inflation models. We obtain
theoretical CMB power spectra as a function of the cosmological
parameters using the CMBFAST and CAMB codes (Seljak \& Zaldarriaga, 
1996; Lewis, Challinor \& Lasenby, 2000). In
order to relate $\sigma_8$ to the CMB power spectrum normalisation, we
first relate $\sigma_8$ to the primordial matter power spectrum
amplitude and then use the analytic expression from Efstathiou, Bond
\& White (1992) to relate this to the $\ell=2$ amplitude of the CMB power
spectrum.  

The COBE data constrain the large scale temperature fluctuations
well, which converts to a strong constraint on $\sigma_8$ for given
values of $h$ and $\Omega_{\rm m}$. The CMB data indicate the position 
of the first acoustic peak, near $\ell\sim 200$ which corresponds to a 
wavenumber of $k \sim 0.03\,\, h \,\,{\rm Mpc}^{-1}$. 
This constrains the combination $\Omega_{\rm m}+\Omega_{\Lambda}$ to be 
roughly around unity (e.g. Efstathiou et al. 1999, Dodelson \& Knox 2000, 
Lange et al. 2000, Balbi et al. 2000, Tegmark and Zaldarriaga 2000), 
consistent with the flat universe assumed in our current analysis. 
In fact using just BOOMERANG and COBE, Lange et al. (2000) find
$\Omega_{\rm m}+\Omega_{\Lambda} \sim 1.1$ (Fig.~2), whereas using just
MAXIMA-1 and COBE, Balbi et al. (2000) find $\Omega_{\rm
m}+\Omega_{\Lambda} \sim 0.9$. At $\sim 1^\circ$ angular scales
the height of the first acoustic peak constrains the matter-radiation
ratio at last scattering, which is proportional to $\Omega_{\rm m}
h^2$. 
In addition, given our assumption of a flat universe, $\Omega_{\rm m}$
and $h$ also significantly affect the position of the first acoustic
peak (see Fig 2. of White, Scott \& Pierpaoli, 2000, for an
illustration). Increasing $\Omega_{\rm m}$ moves the peak to lower $\ell$,
as does increasing $h$. These two effects combine to give the
likelihood distribution in the $\Omega_{\rm m}$-$h$ plane shown in
Fig.~\ref{data} (b). The slightly lower first peak height indicated by
the BOOMERANG and MAXIMA-1 data and the lower $\ell$ position of the
first peak from the BOOMERANG data produce a constraint at higher
$\Omega_{\rm m}$ and $h$ than does the pre-BOOMERANG/MAXIMA-1 compilation
(hereafter pre-BM).  Using the whole compilation together defines 
a region in $(\Omega_{\rm m}$,$h)$ space at the intersection of the 
BOOMERANG$+$MAXIMA-1 and the pre-BM contours. This occurs at high $h$ 
and low $\Omega_{\rm m}$.

\subsection{Type Ia Supernovae} 
\label{sn}

We use the constraints obtained by Perlmutter et al. (1999), which are
fully consistent with those of Riess et al. (1998), based on applying
the classical luminosity-redshift test to distant type Ia
supernovae. The sample consists of 42 high-redshift SN ($0.18 \leq
z\leq0.83$), supplemented by 18 low-redshift SNe ($z < 0.1$). This
analysis determines a combination of $\Omega_{\rm m}$ and
$\Omega_{\Lambda}$. Note that, unlike PV and CMB, SN are insensitive 
to the form of the matter power spectrum and depend only on the overall 
geometry of the universe. 
Since we limit ourselves in this paper to a flat
universe, the SN constraint is translated to a likelihood function of
$\Omega_{\rm m}$, shown in Fig.~\ref{data}~(c).

\section{Comparison and Combination}
\label{results}

\begin{figure*}
\centerline{ 
\begin{tabular}{cc}
\mbox{\epsfig{file=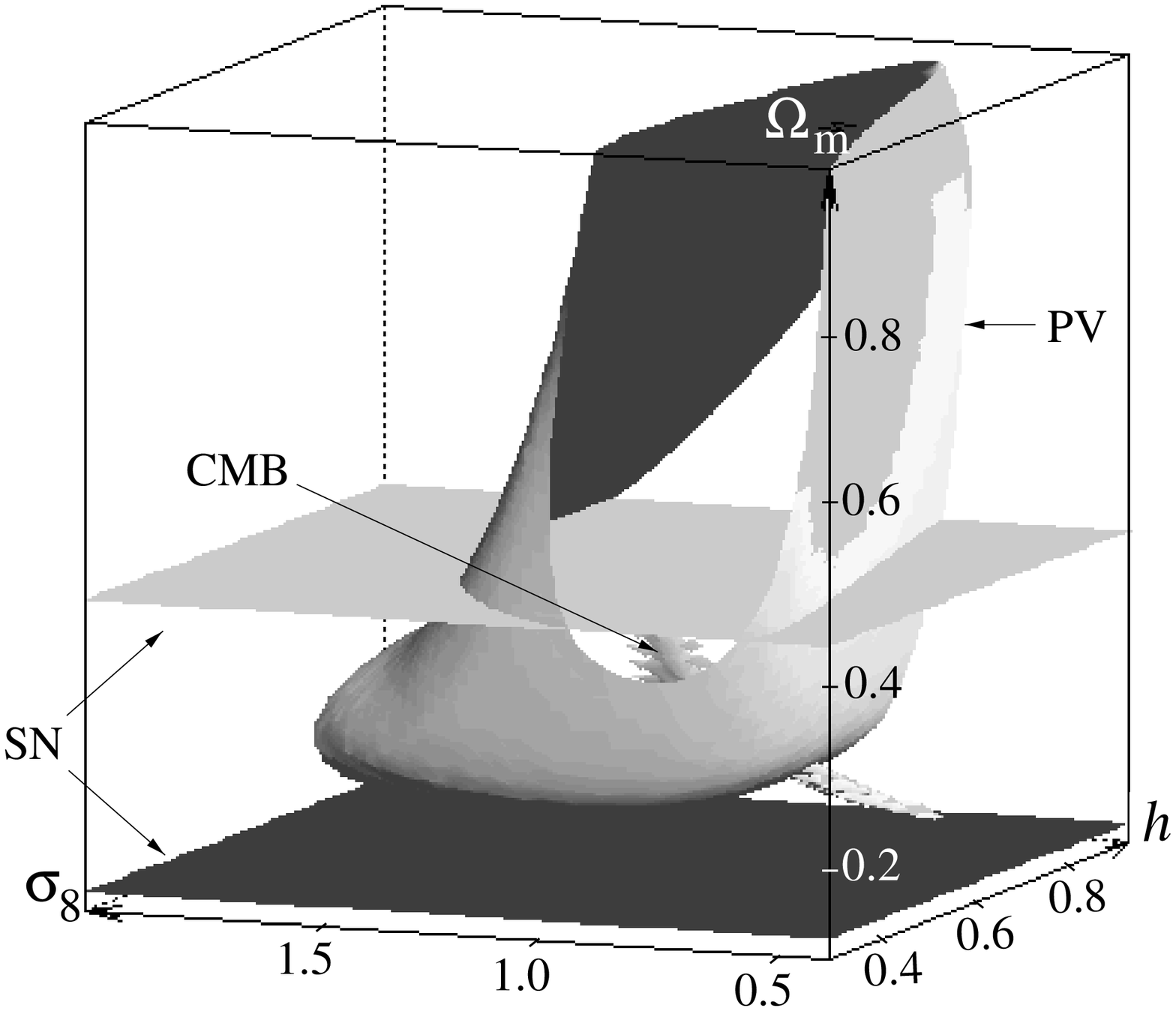,height=8cm}}&
\mbox{\epsfig{file=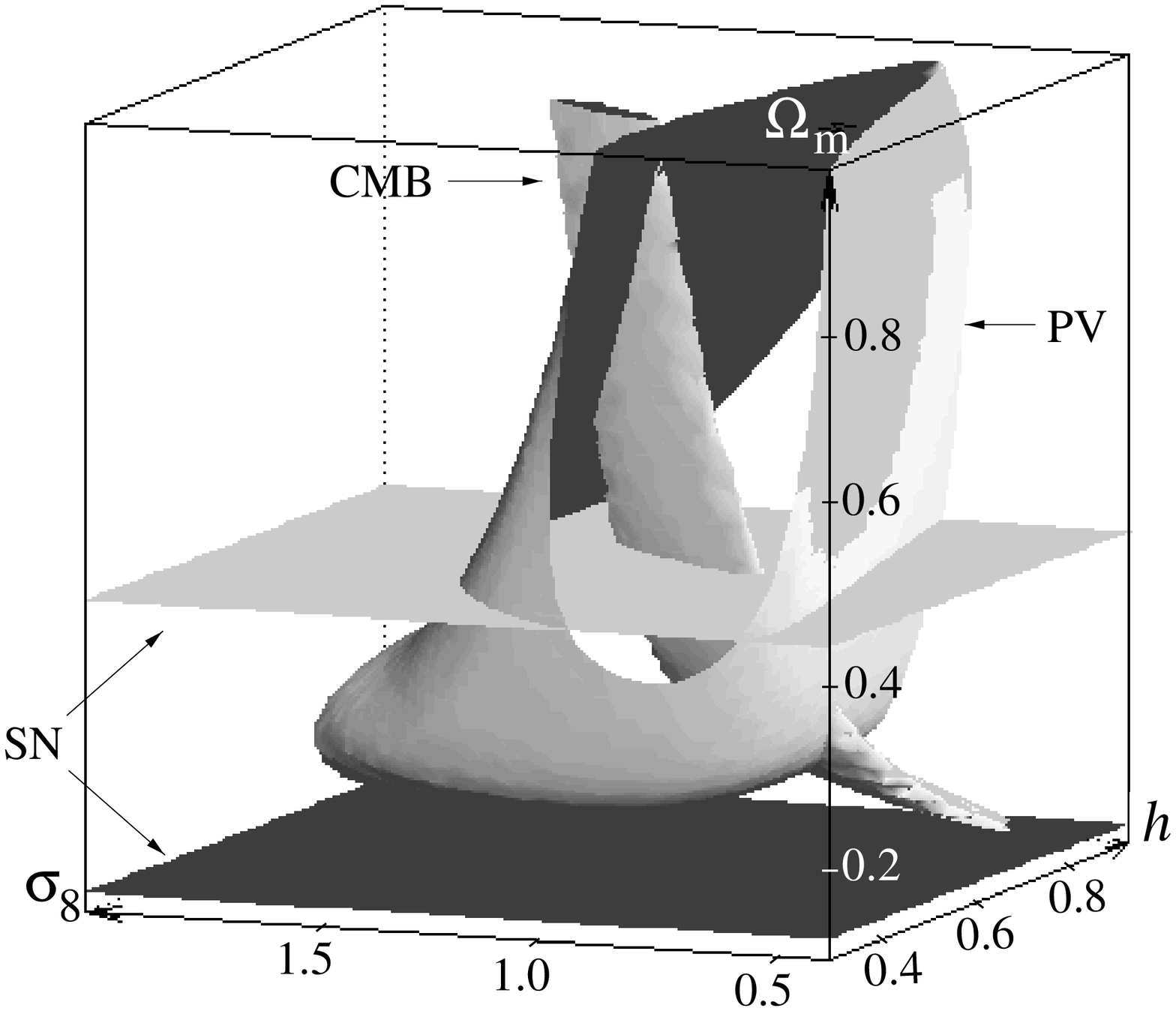,height=8cm}}\\
\mbox{\epsfig{file=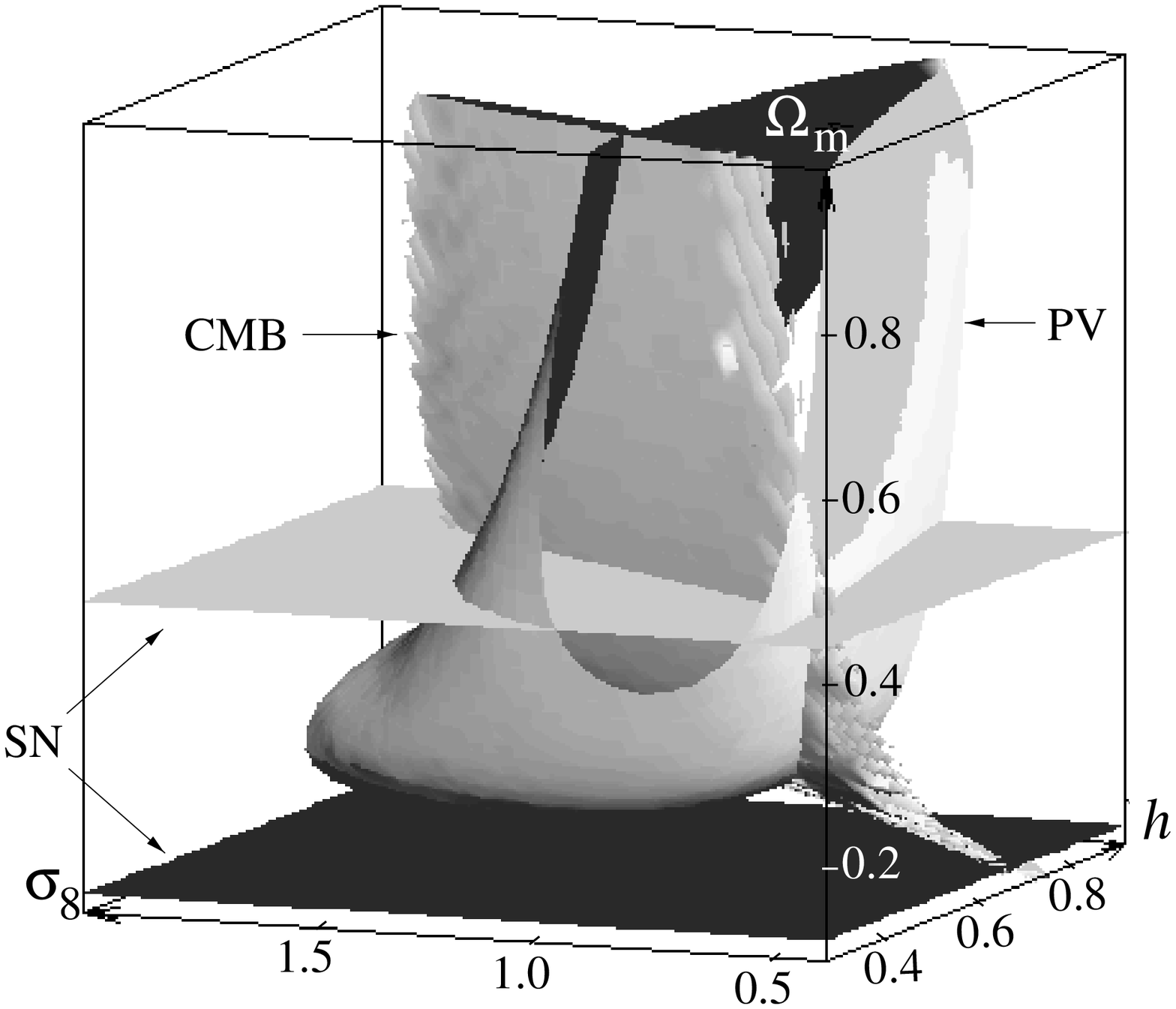,height=8cm}}&
\mbox{\epsfig{file=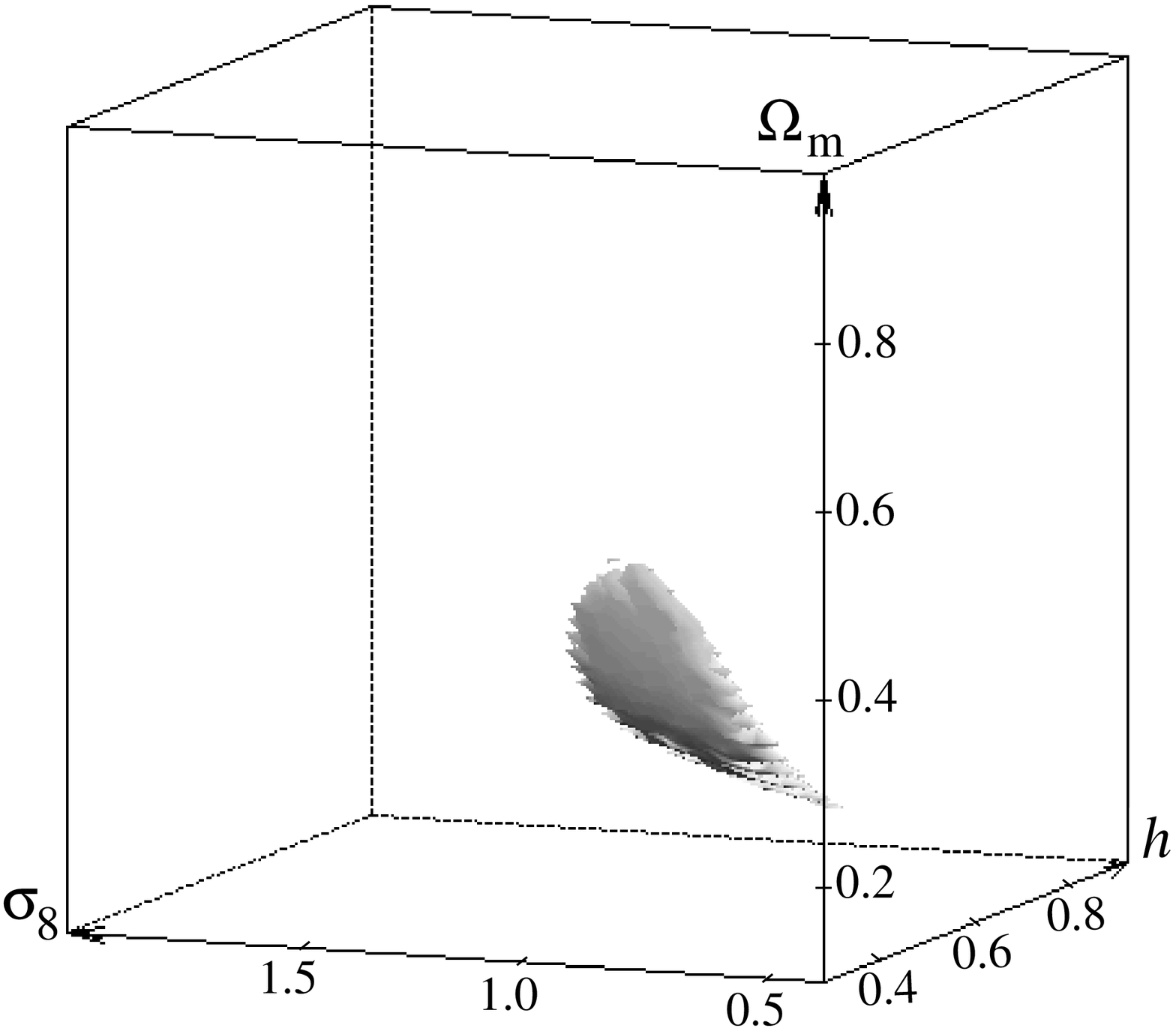,height=8cm}}
\end{tabular}
}
\caption{
Top left:
PV, CMB (whole compilation) and SN 2$\sigma$ iso-probability surfaces.
For PV and CMB the surfaces are at $\Delta$log(Likelihood)=4.01, and
for the SN the surfaces are at $\Delta$log(Likelihood)=2.00,
corresponding to the 95 per cent limits for 3 and 1 dimensional
Gaussian distributions respectively. 
(Integration under the likelihood surfaces can also be used instead of
the likelihood itself in order to define 95 per cent limits, but this
causes little difference to the allowed regions, in the cases shown here.)
The SN surfaces are two horizontal planes.
Top right: the same, but this time the data used for the CMB
surface is just COBE, BOOMERANG (Antarctica flight) and MAXIMA-1.
Bottom left: the same but this time the data used for the CMB
surface is the pre-BM data.
Bottom right: The 2-sigma surface for the joint PV, CMB (whole
compilation) and SN likelihood function.
}
\label{cmbpv3d}
\end{figure*}

In order to examine how well the constraints from PV, CMB and SN agree 
with each other we plot in Fig.~\ref{cmbpv3d} the three corresponding
iso-likelihood surfaces, at the 2-sigma level, in the three-dimensional 
parameter space $(h,\sigma_8,\Omega_{\rm m})$. 
%
%
The upper and lower $95$ per cent limits on $\Omega_{\rm m}$ from SN
are the two horizontal planes. The PV surface encloses a space at
roughly constant $\Omega_{\rm m} h$ and $\sigma_8 \Omega_{\rm
m}^{0.6}$. The CMB surface lies in the intersection of the
regions allowed by each of SN and PV. The fact that the constraints
have a common region of overlap is not trivial; it indicates a
reasonable goodness of fit between the three data sets within the
framework of the assumed cosmological model, which justifies a joint
likelihood analysis aimed at parameter estimation. 
To illustrate the complementary nature of these three data sets we
show in the top right panel of Fig. \ref{cmbpv3d} the same surfaces as
in the top left panel
except that this time the CMB surface was calculated using only the
COBE, BOOMERANG (Antarctica flight) and MAXIMA-1 data. The CMB surface
can be seen to be roughly orthogonal to each of the PV and SN
surfaces. Also for comparison the result of using the pre-BM CMB data
instead is shown in the bottom left panel of Fig. \ref{cmbpv3d}.

Given the very different nature of the three data sets and the
different redshift ranges probed by them ($z\sim 0.02, 0.5, 1000$ for
PV, SN and CMB respectively), we assume that the errors on the
individual data sets are uncorrelated with each other. The likelihood
of a given set of cosmological parameters is thus obtained by
multiplying the three likelihoods of the parameters derived
for each data set alone. The 2-sigma iso-probability surface for the
joint likelihood function is shown in the bottom right panel of
Fig.~\ref{cmbpv3d}. As expected, it is located at the intersection of 
the surfaces from each of the three data sets alone. 
 
The best fit cosmological parameters ($\Omega_{\rm m}$, $h$, $\sigma_8$) 
given all three data sets are given in Table~1, from which we can derive 
$\sigma_8\Omega_{\rm m}^{0.6}=0.54$, $\Omega_{\rm m} h = 0.21$, 
$Q_{\rm{rms-ps}}=19.7 \mu$K and
the age of the universe is $13.2$ Gyr. The CMB power spectrum for this
set of parameters is the solid line plotted in Fig.~\ref{cmbdata},
which can be seen to be a reasonable fit to the data up to the end of the 
first acoustic peak. 
The $\chi^2$ with the CMB data is not simple to quote, since we have
marginalised over the calibration uncertainties for the BOOMERANG and
MAXIMA-1 data. However, using the best fit point to CMB (all
data)+PV+SN, the $\chi^2$ with the pre-BM data
is $52$. This is higher than the number of data points, $39$, which
reflects the fact that the BOOMERANG and MAXIMA-1 points are somewhat
below the other CMB data points. 
Similarly for peculiar velocities, we have marginalised over
$\sigma_{\rm v}$ before calculating the best fit point to PV+CMB+SN, but for
$\sigma_{\rm v}=200$ (the best fit value using peculiar velocities alone)
the $\chi^2$ for the joint best fit point is $1155$, very similar to the
number of data points, $1156$.

\begin{figure*}
\centerline{
\vbox{\epsfig{file=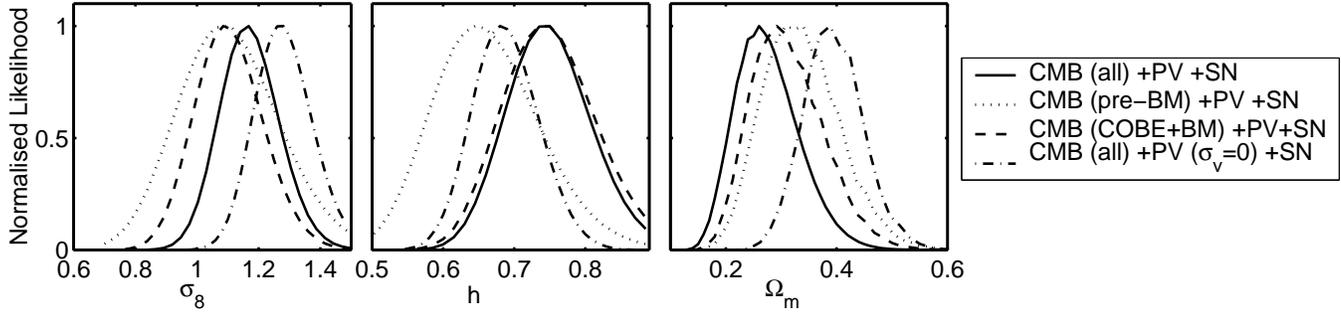,height=18cm, angle=90}}}
\caption
{The 1-dimensional marginalised likelihood distributions 
from the joint PV, CMB and SN likelihood function.
Our main results are shown by the solid lines, which use the whole CMB
data compilation, PV (marginalised over $\sigma_{\rm v}$) and SN. The dotted
lines show the likelihood functions when PV, SN and just the pre-BM
CMB data are used. The dashed line is the result when the CMB data is
just COBE + BOOMERANG + MAXIMA-1. The result (using all the CMB data) when the uncorrelated
velocity dispersion term is not included in the PV analysis
($\sigma_{\rm v}=0$) is shown by the dot-dashed line.
\label{joint}}
\end{figure*}

We may evaluate the probability of a single cosmological parameter,
independent of the values of the other cosmological parameters, by
integrating the probability over the values of the other
parameters. This is what we mean by
`marginalisation'. The solid lines in Fig.~\ref{joint} shows the
resulting 1-dimensional 
marginalised likelihood distributions for each parameter. We obtain
the $95$ per cent limits by integrating the one- dimensional likelihood
distributions and requiring that $95$ per cent of the probability lies
between the quoted limits. These limits are those presented in Table~1. 
The $h$ range agrees well with that from the HST key project of
$h=0.71 \pm 0.06$ (Mould et al. 1999) and the $\Omega_{\rm m}$ limits are
roughly centered on the popular value of $0.3$. 
\begin{table}
\centerline{\vbox{
\begin{tabular}{@{}lcc}
\hline
Parameter & Best fit point & 95 per cent confidence limits\\
\hline
$h$                & $0.74$  &  $0.64 \,<\, h              \,<\, 0.86$ \\
$\Omega_{\rm m}$   & $0.28$  &  $0.17 \,<\, \Omega_{\rm m} \,<\, 0.39$\\
$\sigma_8$         & $1.17$  &  $0.98 \,<\, \sigma_8       \,<\, 1.37$\\
\hline
\end{tabular}
}}
\label{bestfit}
\caption{
Parameter values at the joint PV, CMB, SN optimum.
The $95\%$ confidence limits are given, calculated for each parameter by
marginalising the likelihood function over the other parameters.}
\end{table}

\begin{figure*}
\centerline{
\vbox{\epsfig{file=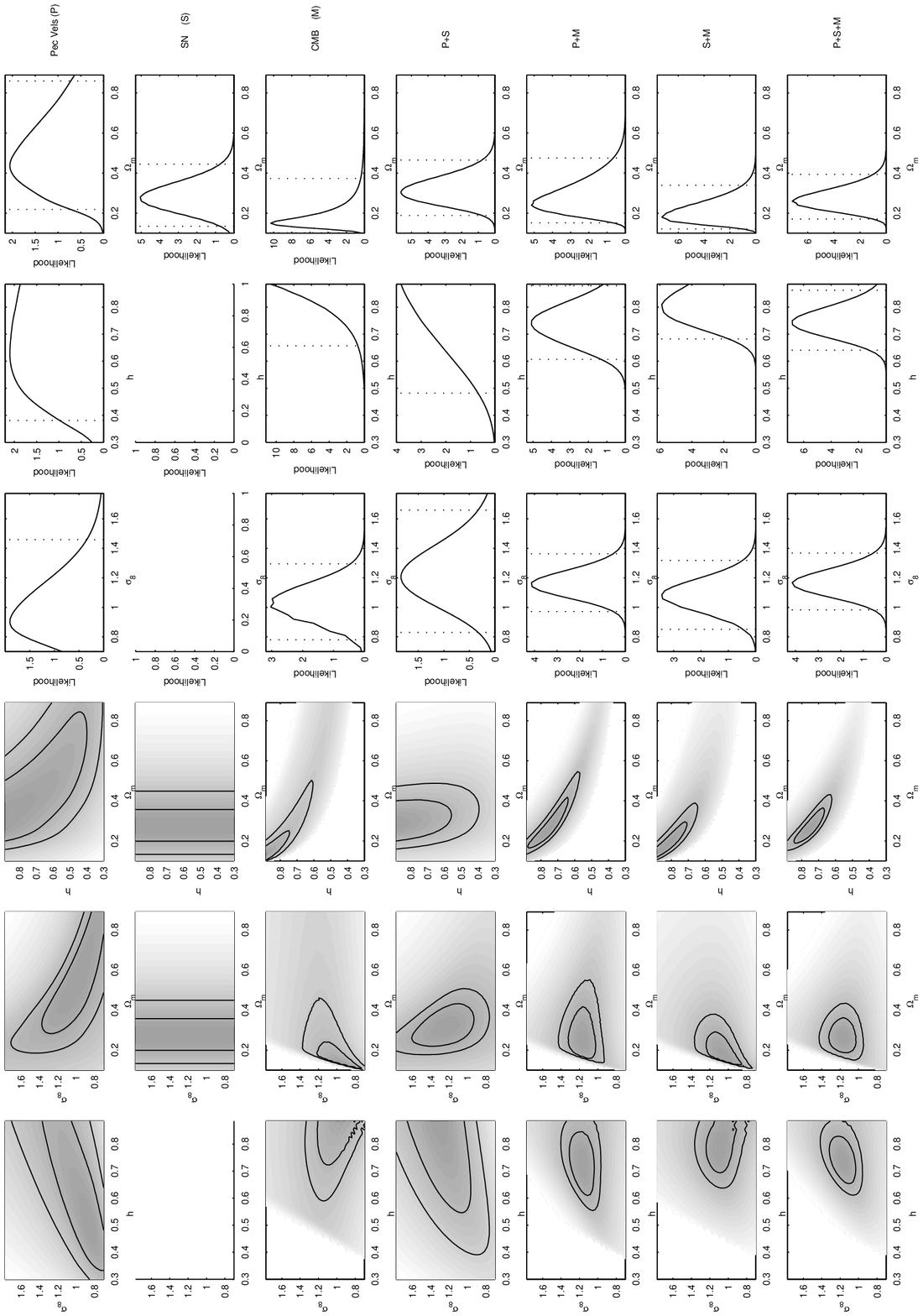,height=22cm,
angle=180}}}
\caption
{
Likelihood functions for each individual data set (top three rows:
left hand side of this page), for the combinations of pairs of data
sets (next three rows, marked P+S, P+M and S+M), and for the
combination of all three data sets (bottom row: right hand side of the
page, marked P+S+M). Columns 1 to 3 (bottom of page) show the 2-dimensional 
marginalised likelihood functions, with solid lines at the 68 and 95 per cent 
confidence limits, in each case the third parameter has 
been marginalised over. Columns 4 to 6 (top of page) are the 1-dimensional 
marginalised distributions for each cosmological parameter,
with dotted lines showing the 95 per cent confidence regions. Note that the 
whole CMB data compilation was included in this analysis, and the PV term 
$\sigma_{\rm v}$ was marginalised over.
\label{all}}
\end{figure*}

Fig.~\ref{all} shows together the 2-dimensional and 1-dimensional
marginalised distributions as evaluated for each data set alone and
then jointly for each pair of data sets, and finally for the three
data sets together (again, the 68 and 95 per cent limits are found by
integration of the probability distributions). For the pairs of data
sets, or the single data sets alone, there is some dependence of the
confidence regions on the ranges used in the marginalisation. However,
the results when all three data sets are used are insensitive to the
ranges of integration we have used, except for the limit of $h<0.9$, 
which we consider to be a reasonable prior. 
A measure of the excellent agreement between these data sets is given
by the similarity of the parameter constraints from the three
different possible pairings of the data sets (1d plots marked P+S, P+M
and S+M in Fig. \ref{all}). Note also that the CMB data alone
prefer a high $h$, but on combining with PV and SN there is an upper
bound which is just below our prior of $h<0.90$. A detailed
examination in 3-dimensions reveals that inclusion of the PVs cuts off
a high $h$, low $\Omega_{\rm m}$ part of the CMB surface, and
inclusion of SN cuts off a high $h$, high $\Omega_{\rm m}$  part of
the CMB surface, thus lowering the preferred value of $h$. 

We have repeated the entire analysis using different subsets of CMB
data. Using the pre-BM data the 1-dimensional
marginalised likelihood functions (dotted lines in Fig. \ref{joint})
are in good agreement but somewhat wider than when using all the data,
especially in the constraint on $h$ which extends to lower values than
before. Using just the COBE, BOOMERANG (Antarctica flight) and
MAXIMA-1 data (dashed lines) the results are very similar to when all
data is used. 
At first this may seem surprising given the much larger 3-dimensional
surface (Top right versus top left panels of Fig. \ref{cmbpv3d} and
Fig. \ref{data} (b)) but the high $\Omega_{\rm m}$, $h$ part is ruled
out by both PV and SN, leaving virtually the same region as when all
CMB data are used. 

In the region of the power spectrum where a second acoustic peak is
predicted, we note that our best fitting models are not a
good fit to the data, producing more power than observed by both
BOOMERANG and MAXIMA-1. The easiest way to reconcile this is to
increase $\Omega_{\rm b} h^2$ to a value approximately double 
that found from nucleosynthesis (eg. Hu 2000, White et al. 2000). 
This has the effect of increasing the
heights of the odd numbered peaks and decreasing those of the even
numbered peaks. It also has the effect of reducing the sound horizon
and thus shifting the first peak to even smaller angular scales. 
Repeating some of our analysis using $\Omega_{\rm b} h^2=0.04$
(and assuming that the PV likelihood function is relatively
insensitive to this value) we find that the agreement between the data
sets is still good, and the constraints on $\sigma_8$ and $\Omega_{\rm
m}$ are not significantly affected. However, the the best-fitting $h$
value tends towards our upper limit of $0.90$, which allows the peak
to be at smaller angular scales (White et al. 2000, Fig 2).

In order to check the level of sensitivity to the use of $\sigma_{\rm v}$ 
in the PV analysis we have also repeated the joint analysis using peculiar 
velocity likelihoods that were obtained without including this additional 
term (Section \ref{pv}).  This is the linear method used  
in the main section of Freudling et al. (1999) and in Zehavi \& Dekel 
(1999). As mentioned already, the linear analysis prefers slightly 
higher values of $\sigma_8 \Omega_{\rm m}^{0.6}$. As a result,
the region of agreement between PVs, CMB and SN now only just occurs  
at the 2-sigma level.
The resulting 1-dimensional marginalised likelihood functions are the
dot-dashed lines in Fig. \ref{joint}.
They are not very different from those obtained with the non-linear 
correction, having slightly higher $\Omega_{\rm m}$ and $\sigma_8$ and 
a slightly lower $h$. 
It is encouraging to see that the joint analysis results are 
fairly robust to the uncertainties in the PV analysis.

We also quote the results for our main analysis (whole CMB data
compilation, marginalised over $\sigma_{\rm v}$) in terms of the natural 
parameter combinations for PVs in Table~2. The range in $\sigma_8 \Omega_{\rm
m}^{0.6}$ is slightly lower than that preferred by peculiar velocities
alone, mainly due to the orthogonal constraint from the CMB on
$\sigma_8$ and $\Omega_{\rm m}$ in the $\Omega_{\rm m}$ range allowed
by SN, which disfavours larger $\sigma_8 \Omega_{\rm m}^{0.6}$. 
The $\Omega_{\rm m} h$ limits are much tighter and at the low
end of those provided by peculiar velocities alone, again mainly due to
the CMB constraint in the range allowed by SN.
\begin{table}
\centerline{\vbox{
\begin{tabular}{@{}lcc}
\hline
Parameter & Best fit point & 95 per cent confidence limits\\
\hline
$\sigma_8 \Omega_{\rm m}^{0.6}$    
  & $0.54$ & $0.40 \,<\, \sigma_8 \Omega_{\rm m}^{0.6} \,<\, 0.73$\\
$\Omega_{\rm m} h$                  
  & $0.23$ & $0.16 \,<\, \Omega_{\rm m} h              \,<\, 0.27$ \\
$\Omega_{\rm m}$           
  & $0.28$ & $0.18 \,<\, \Omega_{\rm m}                \,<\, 0.42$\\
\hline
\end{tabular}
}}
\label{bestfit_so_oh}
\caption{
Parameter values at the joint PV, CMB, SN optimum and $95$ per cent
limits in terms of the parameters $\sigma_8 \Omega_{\rm m}^{0.6}$,
$\Omega_{\rm m} h$ and $\Omega_{\rm m}$. 
} 
\end{table}

\section{Conclusion}
\label{conclusion}

We have performed a joint analysis of three complementary data sets
free of galaxy-density biasing, using peculiar velocities, CMB
anisotropies, and high-redshift supernovae. The constraints from the
three data sets overlap well at the 2-sigma level and there is acceptable
goodness of fit. These data sets constrain roughly orthogonal
combinations of the cosmological parameters, and are combined to
provide tighter constraints on the parameters (Table~1). These
constraints are found to be reasonably robust to the CMB data compilation 
used, the peculiar velocity catalogue used, and the assumption of an 
uncorrelated velocity dispersion at zero lag (Fig.~\ref{joint}).

The values obtained from the joint analysis for $h$ and $\Omega_{\rm m}$, 
and for the combinations of cosmological parameters (Table~2), 
are in general agreement with other estimates (eg. Bahcall et al. 1999), 
but this analysis tends to favor a slightly higher value for $\sigma_8$. 
In particular, the result for $\sigma_8$ is higher than the Bridle et al. 
(1999) constraint, $\sigma_8 = 0.74 \pm 0.1$ ($95\%$ confidence) 
obtained by combining the CMB with cluster abundance and IRAS and allowing 
for linear biasing. This may reflect the preference of the peculiar 
velocities for a slightly higher value of $\sigma_8 \Omega_{\rm m}^{0.6}$ 
than favored by the cluster abundance analysis. The implications of 
considering the constraints arising from all the above mentioned probes 
will be discussed elsewhere.  

The addition of BOOMERANG and MAXIMA-1 to our CMB data compilation
brought down the height of the first acoustic peak and shifted it to 
larger angular scales, which both increase a combination of $h$ and
$\Omega_{\rm m}$. The combination of BOOMERANG and MAXIMA-1 with the 
older CMB data had the effect of breaking the degeneracy between $h$ and
$\Omega_m$ and leaving a high $h$ region of parameter space.
The resulting constraint on the Hubble constant, 
$h=0.75 \pm 0.11$ ($95$ per cent confidence), 
agrees well with that from the HST key project value of $h=0.71 \pm 0.06$.
This result is also similar to that of Lange et al. (2000, Table~1, P10). 

Note that in this analysis we take all the data sets used at equal
weight. An extension to this work would be to allow freedom in the
weights given to the different probes, as in Lahav et al. (2000).

\subsection*{ACKNOWLEDGMENTS}

We thank Gra\c ca Rocha for her work on compilation of the CMB data
set and George Efstathiou for providing the supernova likelihoods. 
SLB acknowledges the PPARC for support in the form of a research
studentship.
This work was supported by the DOE and the NASA grant NAG 5-7092
at Fermilab,  by US-Israel Binational Science Foundation grants
95-00330 and 98-00217, and by Israel Science Foundation grant 546/98.
SLB and IZ acknowledge the hospitality of the Hebrew University of 
Jerusalem. 

\bibliographystyle{/opt/TeX/tex/bib/mn}

\bsp 
\label{lastpage}

\end{document}